# High-resolution population estimation using household survey data and building footprints

Gianluca Boo[1*], Edith Darin[1], Douglas R Leasure[1], Claire A Dooley[1], Heather R Chamberlain[1], Attila N Lázár[1], Kevin Tschirhart[2], Cyrus Sinai[3,4], Nicole A Hoff[3], Trevon Fuller[3], Kamy Musene[3], Arly Batumbo[5], Anne W Rimoin[3], Andrew J Tatem[1]

[1]WorldPop, School of Geography and Environmental Science, University of Southampton, Southampton, UK

[2]Center for International Earth Science Information Network (CIESIN), Columbia University, New York, NY, USA

[3]UCLA Fielding School of Public Health, University of California at Los Angeles, Los Angeles, CA, USA

[4]Department of Geography, University of North Carolina at Chapel Hill, Chapel Hill, NC, USA

[5]Bureau Central du Recensement, Institut National de la Statistique, Kinshasa, Democratic Republic of the Congo

*Corresponding author — gianluca.boo@gmail.com

## Abstract

The national census is an essential data source to support decision-making in many areas of public interest. However, this data may become outdated during the intercensal period, which can stretch up to several decades. We developed a Bayesian hierarchical model leveraging recent household surveys with probabilistic sampling designs and building footprints to produce up-to-date population estimates. We estimated population totals and age and sex breakdowns with associated uncertainty measures within grid cells of approximately 100m in five provinces of the Democratic Republic of the Congo, a country where the last census was completed in 1984. The model exhibited a very good fit, with an $R^2$ value of 0.79 for out-of-sample predictions of population totals at the microcensus-cluster level and 1.00 for age and sex proportions at the province level. The results confirm the benefits of combining household surveys and building footprints for high-resolution population estimation in countries with outdated censuses.

# Introduction

Accurate population figures are essential to support decision-making in many areas of public interest, for instance, urban planning, environmental hazard risk management, and public health[1]. To this end, the most complete and reliable data source is arguably the national population and housing census[1,2]. However, the data collected in the census may be incomplete due to inaccessible regions and may quickly become outdated because of migration, fertility, and mortality patterns occurring during the ten-year intercensal period, which can occasionally stretch up to several decades[3]. In such circumstances, the census data can be completed using different population estimation techniques[1-4]. Among these techniques, the "bottom-up" population modeling approach offers the advantage of producing up-to-date population estimates independently from the national census[3,4].

Bottom-up models leverage population data retrieved from recent household surveys involving the complete enumeration of a representative sample of small and well-defined areas, named microcensus clusters[3]. In their basic form, these models link cluster-level population totals and ancillary geospatial covariates with complete coverage of the region of interest, such as settlement extents[5,6] and satellite imagery classes[7,8], to estimate population totals in unsurveyed areas. These models can also include additional geospatial covariates[5-8], administrative or functional strata[5,6], and existing age and sex structures to disaggregate the population estimates within different age and sex groups[9]. The United Nations Population Fund recently highlighted the role of bottom-up population models for census planning and preparation[4].

In this study, we extended an existing Bayesian hierarchical framework for bottom-up population modeling recently applied in Nigeria[6]. We integrated a weighted-precision approach to produce unbiased estimates from household surveys with population-weighted sampling designs[10,11] and modeled age and sex structures within administrative provinces using the same household survey data[12]. We also accessed building footprints to provide an accurate approximation of the settled area and derive morphological and topological attributes incorporated in the model both as geospatial covariates[13,14] and functional strata representing different settlement types[14,15]. Our model was designed to estimate population totals and age and sex breakdowns together with associated uncertainty measures within grid cells of 3 arc seconds, approximately 100m. These high-resolution population estimates can be flexibly aggregated within different geographic units to support specific use cases[16].

We applied our bottom-up population model in five provinces in the western part of the Democratic Republic of the Congo (DRC). In this country, the last national census was completed in 1984, and existing projections are considered erratic because of the significant security, economic, and institutional tribulations of the past decades[17]. The uncertainty around the geography and demography of the Congolese population is particularly critical to decision-making for public health and humanitarian interventions[17,18]. Our modeling effort provides up-to-date population estimates at high spatial resolution in five provinces of the DRC while advancing the state-of-the-art of bottom-up population modeling in the use of survey data with widely adopted probabilistic sampling designs[1,2] and building footprint attributes that are, in part, openly available across sub-Saharan Africa[19].

## Results

### Population estimates

We developed a hierarchical Bayesian model to estimate population totals and age and sex breakdowns at high spatial resolution in five provinces in the western part of the DRC. **Figure 1** shows the estimated population totals within grid cells of approximately 100m for the five provincial capitals. Kinshasa (**Figure 1A**) has the most considerable spatial extent and the highest population totals per grid cell, with large variations between the central part of the city and its outskirts. The remaining cities (**Figure 1B-E**) have reduced extents and lower population totals per grid cell, confirming the predominantly rural character of the study region. These high-resolution population estimates, including totals, age and sex breakdowns, and associated measures of model uncertainty, can be accessed on a dedicated web repository[20], visualized using a web-mapping application[21], and integrated into data analyses through an R package[22].

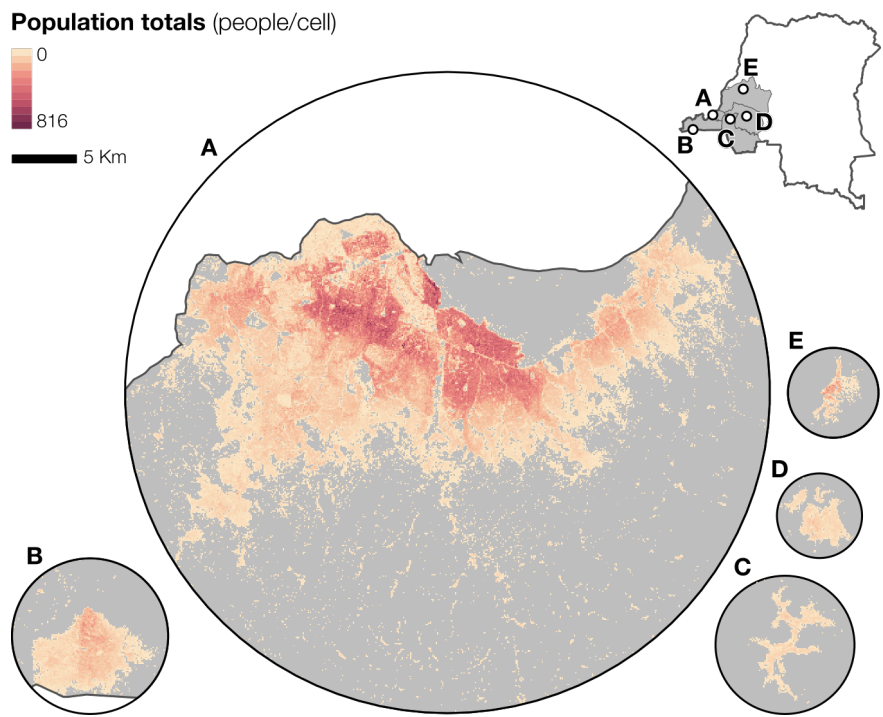

**Fig 1.** Estimated population totals per approximately 100m grid cell in the capital cities of the provinces of (**A**) Kinshasa (Kinshasa), (**B**) Kongo Central (Matadi), (**C**) Kwango (Kenge), (**D**) Kwilu (Bandundu), and (**E**) Mai-Ndombe (Inongo). The map of the DRC shows the extent of the five provinces in grey.

## Population totals and densities

As presented in Eq. (**1**), we modeled population totals as a Poisson process resulting from estimated population densities (i.e., people/total area of building footprints) multiplied by the total area of building footprints within the microcensus clusters. We computed population totals and densities across 926 clusters based on data collected across two rounds of household surveys and the total area of the building footprints. We discarded 21 clusters exhibiting spurious population densities because 7 clusters had reduced survey coverage due to inaccessible areas and 14 clusters had no building footprints coverage. **Figure 2** shows the geographic distribution of the observed population densities (people/building footprint ha) across the 905 microcensus clusters according to the settlement type and province. The clusters in the provinces of Kwango (**Figure 2C**), Kwilu (**Figure 2D**), Mai-Ndombe (**Figure 2E**), the eastern part of Kinshasa (**Figure 2A**), and Kongo Central (**Figure 2B**) were primarily rural, with highly heterogeneous population densities. Most urban clusters were located in the provinces of Kinshasa and Kongo Central, the most urbanized part of the study region, where population densities were generally homogeneous and lower than rural clusters.

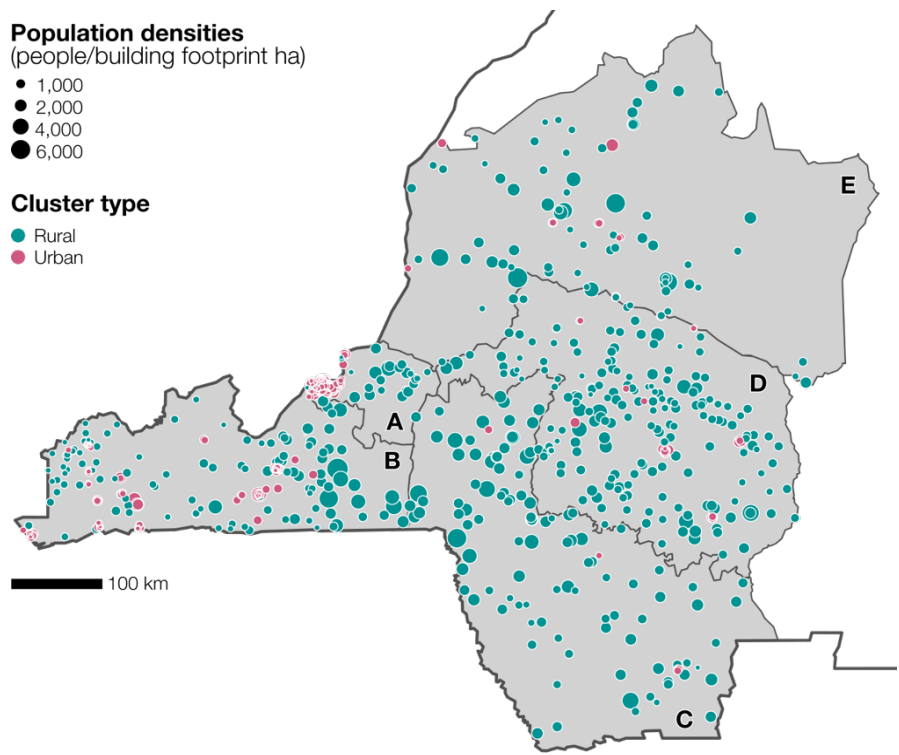

**Fig 2.** Observed population densities (people/building footprint ha) in the 905 microcensus clusters according to settlement type for the provinces of (**A**) Kinshasa — 25 rural and 254 urban clusters, (**B**) Kongo Central — 113 rural and 108 urban clusters, (**C**) Kwango — 110 rural and 6 urban clusters, (**D**) Kwilu — 183 rural and 23 urban clusters, and (**E**) Mai-Ndombe — 69 rural and 12 urban clusters.

**Hierarchical intercepts**

As presented in Eq. (**2**), we defined population log-densities in the 905 microcensus clusters as the response variable of a linear regression. We estimated the random intercept hierarchically by settlement type (n=2), province (n=5), and sub-provincial region (n=37) (Eq. (**8**)). **Figure 3** shows the posterior probability distributions of the hierarchical intercept by settlement type and province. While the distributions were generally similar across rural and urban settlements, the posterior means were lower in the urban settlements of the Kinshasa (**Figure 3A**) and Kongo Central (**Figure 3B**) provinces, potentially because of the higher prevalence of non-residential buildings. As a consequence of the limited coverage and the heterogeneous residential context, the 95% credible intervals are wider in urban settlements, especially in the provinces of Kongo Central (**Figure 3B**), Kwango (**Figure 3C**), and particularly Mai-Ndombe (**Figure 3E**).

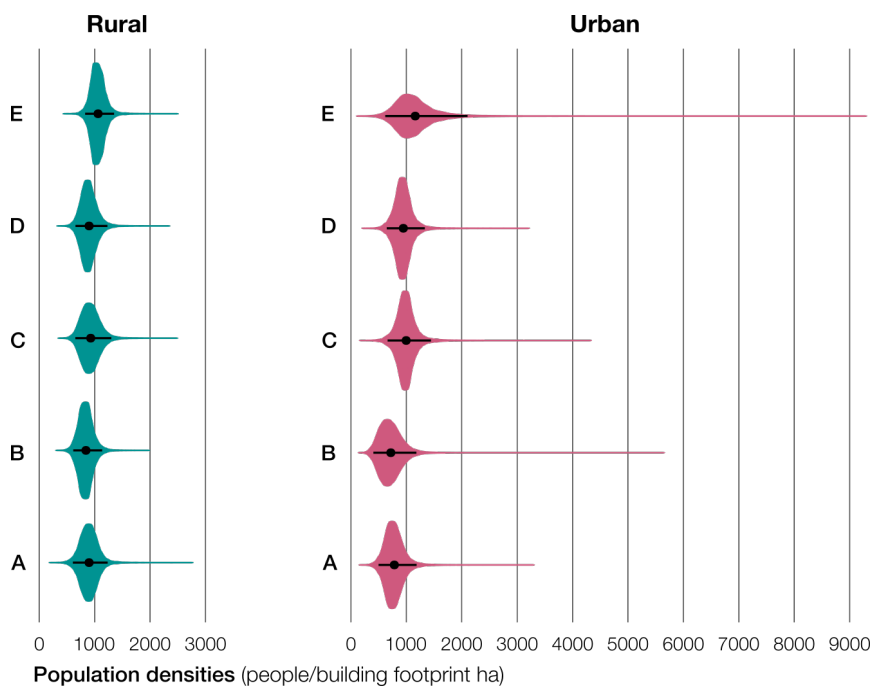

**Fig 3.** Posterior probability distributions of the random intercepts (people/building footprint ha) by settlement type across the provinces of (**A**) Kinshasa, (**B**) Kongo Central, (**C**) Kwango, (**D**) Kwilu, and (**E**) Mai-Ndombe. The black dots show the mean of the distributions, while the horizontal black lines show the 95% credible intervals.

## Covariate effects

As presented in Eq. (**7**), we combined the hierarchical intercepts with the additive effects of three geospatial covariates derived from building footprint attribute summaries. We first defined the covariate effects to be independently estimated by settlement type for each covariate (Eq. (**9**)) and, if their posterior distribution was similar across rural and urban settlements, we converted them into fixed effects (Eq. (**10**)). **Figure 4** shows the covariate effects estimated in the model; we estimated random effects by settlement type for the first two covariates and a fixed effect for the third covariate. While the covariate *Average Building Proximity* (i.e., the inverse of the average distance to the nearest building footprint) had a significant positive effect at the 95% credible level in rural settlements, the effect was non-significant in urban settlements. Conversely, the covariate *Average Building Focal Count* (i.e., the average count of building footprints in a focal window of approximately 2km) had a significant negative effect at the 95% credible level in rural settlements and a significant positive effect in urban settlements. The covariate *Average Building Area* (i.e., the average area of the building footprints) had a strong significant negative effect at the 95% level across both settlement types and was converted into a fixed effect.

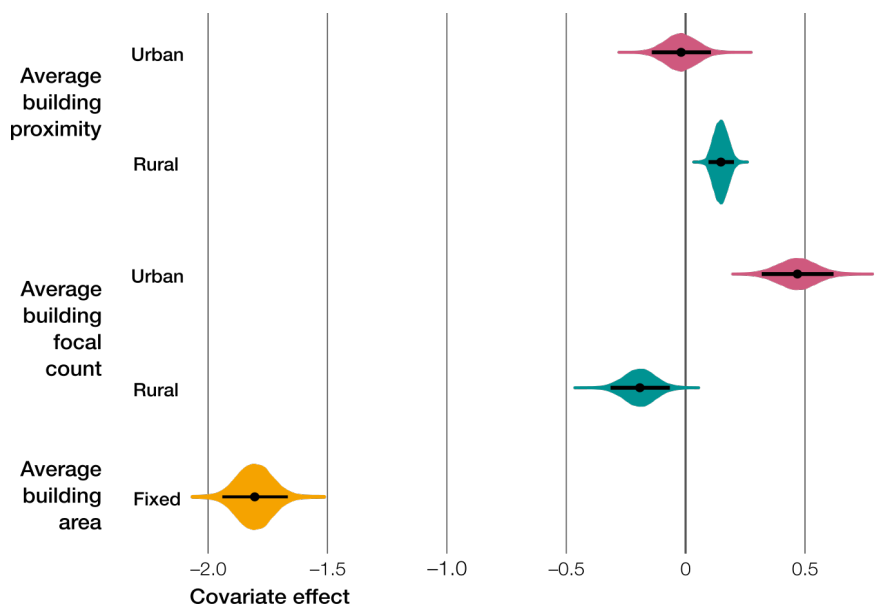

**Fig 4.** Posterior probability distribution of the random effect by settlement type (i.e., urban and rural) for the covariates *Average Building Proximity* and *Average Building Focal Count* and the fixed effect for the covariate *Average Building Area*. The black dots show the mean of the distributions, while the horizontal black lines show the 95% credible intervals.

**Age and sex proportions**

As presented in Eq. (**11, 12**), we modeled age and sex proportions from the household survey data aggregated at the province level with a Dirichlet-multinomial process. **Figure 5** shows the means of the posterior distribution of the age and sex proportions with relative 95% credible intervals for the five provinces. Age and sex structures were similar in the predominantly rural provinces of Kwango (**Figure 5C**), Kwilu (**Figure 5D**), and Mai-Ndombe (**Figure 5E**). In these provinces, the bases of the pyramids were large and became increasingly narrow for older age groups. In the predominantly urban provinces of Kinshasa (**Figure 5A**) and Kongo Central (**Figure 5B**), the pyramids had a narrower base, typically associated with lower fertility. The province of Kinshasa also had a larger part of the population between 20 and 49 years old, potentially because of work-related migratory patterns. The 95% credible intervals were generally narrow because of the limited differences in the aggregated province-level age and sex structures, which were not reflective of potential patterns occurring at the sub-provincial level.

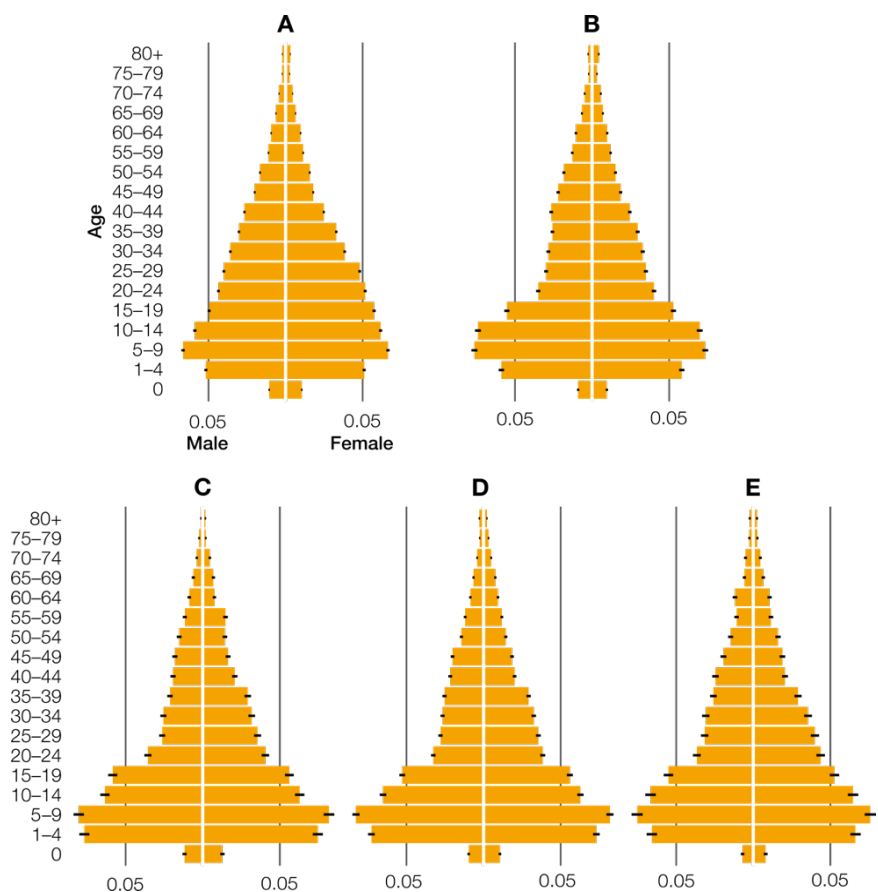

**Fig 5.** Means of the posterior distribution of age and sex proportions for the provinces of (**A**) Kinshasa, (**B**) Kongo Central, (**C**) Kwango, (**D**) Kwilu, and (**E**) Mai-Ndombe. The horizontal black lines show the 95% credible intervals.

**Model diagnostics**

We achieved model convergence in 10,000 sample iterations for the three MCMC chains. **Table 1** summarizes the analysis of residuals for population totals (people) and population densities (people/building footprint ha) at the microcensus-cluster level and age and sex proportions at the cluster level for in-sample and out-of-sample posterior predictions. The analysis suggested a very good model fit for population totals for in-sample ($R^2 = 0.81$) and out-of-sample ($R^2 = 0.79$) predictions, despite a limited fit for population densities for in-sample ($R^2 = 0.52$) and out-of-sample ($R^2 = 0.47$) predictions. For both population totals and densities, approximately 90% of the observations were within the 95% credible intervals of out-of-sample predictions, suggesting that the uncertainty intervals were robust. The analysis also indicated slight over-prediction of population totals and slight under-prediction of population densities at the microcensus-cluster level, with larger imprecision and inaccuracy for the latter. Province-level age and sex proportions had a perfect fit for in-sample ($R^2 = 1.00$) and out-of-sample ($R^2 = 1.00$) model predictions, with imperceptible levels of imprecision and inaccuracy. In both cases, 100% of the observations fell within the 95% credible intervals, suggesting conservative uncertainty intervals for age and sex proportions estimated at the province level. However, these credible intervals are unlikely to account for sub-provincial differences in age and sex proportions.

**Table 1.** Analysis of residuals for the estimated population totals (people), population densities (people/building footprint ha), and province-level age and sex proportions for in-sample and out-of-sample posterior predictions. Bias represents the mean of the residuals, imprecision the standard deviation of residuals, inaccuracy the mean of absolute residuals, $R^2$ the squared Pearson correlation coefficient among the residuals, and the percentage of observations falling within the 95% credible intervals. Values in parentheses are computed using scaled residuals (residuals/predictions).

| Estimate | Prediction | Bias | Imprecision | Inaccuracy | $R^2$ | 95% CI |
|---|---|---|---|---|---|---|
| **Population totals** | In-sample | 13.81 (–0.02) | 165.50 (0.41) | 100.17 (0.27) | 0.81 | 92.49% |
| **Population totals** | Out-of-sample | 13.43 (–0.03) | 173.28 (0.44) | 105.61 (0.29) | 0.79 | 90.50% |
| **Population densities** | In-sample | –13.96 (–0.02) | 441.66 (0.41) | 266.32 (0.27) | 0.52 | 92.04% |
| **Population densities** | Out-of-sample | −15.20 (−0.03) | 464.69 (0.44) | 282.90 (0.29) | 0.47 | 90.06% |
| **Age and sex proportions** | In-sample | 0.00 (0.00) | 0.00 (0.00) | 0.00 (0.00) | 1.00 | 100.00% |
| **Age and sex proportions** | Out-of-sample | 0.00 (0.00) | 0.00 (0.00) | 0.00 (0.00) | 1.00 | 100.00% |

**Figure 6** visually contrasts the observed population totals (people) and densities (people/building footprint ha) versus in-sample and out-of-sample posterior predictions according to settlement type. The figure confirms a very good model fit for population totals, which were generally in line with the observed totals. The most populated microcensus clusters were located in urban settlements, where 93.00% and 91.22% observations fell within the 95% credible intervals of in-sample and out-of-sample posterior predictions. The limited model fit for population densities was due to large underpredictions in densely populated clusters located in rural settlements. These clusters were characterized by a reduced coverage of the building footprints, which artificially increased the observed population densities, where 91.58% and 90.82% observations fell within the 95% credible intervals of in-sample and out-of-sample posterior predictions. The impact of the population density outliers was partly corrected when estimating population totals because of the use of the building footprint area as a multiplicative constraint.

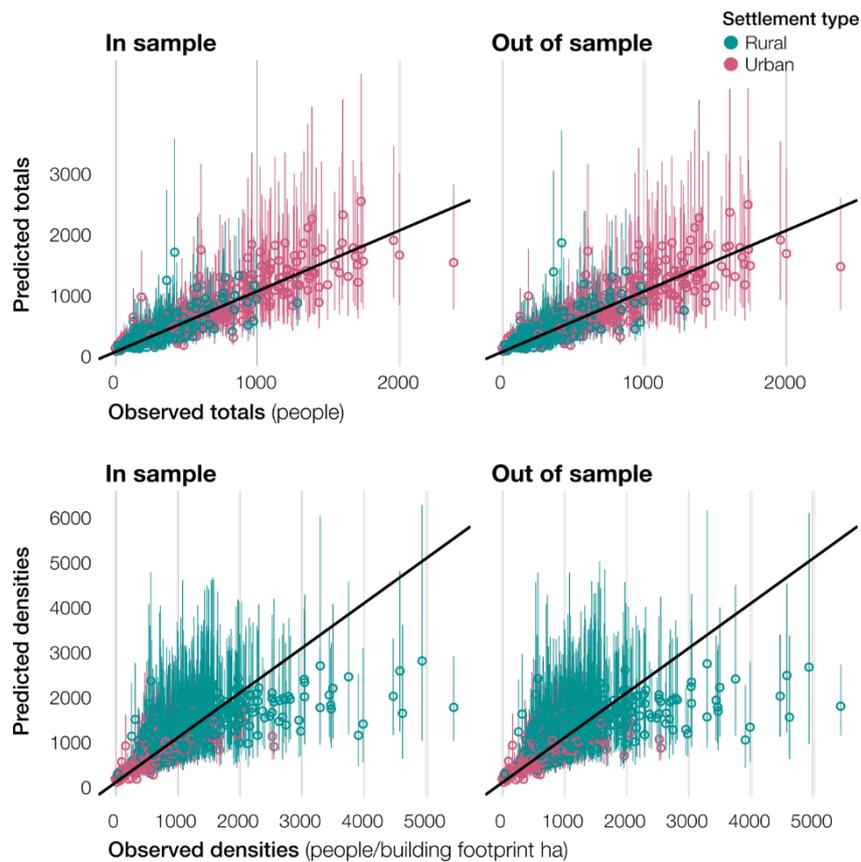

**Fig 6.** Observed microcensus-cluster level totals (people) and densities (people/building footprint ha) versus in-sample and out-of-sample mean posterior predictions (dots) with 95% credible intervals (vertical lines). Population totals and densities are classified by settlement type (i.e., urban and rural). The diagonal black lines show a perfect relationship between observations and predictions.

## Discussion

In this study, we extended a Bayesian hierarchical framework for bottom-up population modeling to leverage household survey data with different sampling designs and building footprint attributes. Existing techniques for population estimation demonstrated the advantages of using household surveys in terms of time and cost to fully enumerate a set of representative clusters compared with the country-wide coverage of a national census[1,2]. While these estimation techniques are generally endorsed for decision-making[1,2] and census support[8], the resulting estimates should not be seen as a substitute replacement for the richness of information collected in the census. However, when the census data is incomplete or outdated, bottom-up modeling is an effective approach to produce comprehensive and up-to-date population estimates at high resolution[3,4]. These estimates can be flexibly aggregated within different units, such as administrative boundaries, catchment areas, health districts, or custom-made polygons, to support different applications[16].

Our modeling effort estimated population totals, age and sex breakdowns at a spatial resolution of approximately 100m together with measures of model uncertainty. The model leveraged household survey data with a population-weighted sampling design[23], typically adopted in national household surveys, such as Demographic and Health Surveys (DHS)[11]. Given that this type of design oversamples locations with higher population densities, we included a weighted-precision approach to recover unbiased estimates of population totals and densities with robust credible intervals[11]. This approach is often regarded as difficult to implement[10], as we confirmed through a systematic analysis of the weights used in the population-weighted sampling[24]. Our assessment exposed the presence of weight outliers associated with uncertainties in the population data used to compute the sampling weights that we truncated at the 90$^{th}$ percentile of the statistical distribution. The use of sampling weights associated with the seeds of the clusters also introduces additional uncertainty in our weighted-precision approach because it does not accurately represent the probability of selection of the entire cluster.

We accessed building footprints to provide an accurate approximation of the settled area and derive morphological and topological attributes from building footprints to produce model covariates and define hierarchical intercepts and random effects by settlement type. These attributes are, in part, openly available across sub-Saharan Africa[15,19]. The hierarchical intercept by settlement type and province suggested higher population densities (people/building footprint ha) in rural than urban settlements. This result was confirmed by the observed population densities at the microcensus-cluster level, where we discarded 14 clusters with no building footprint. Limited building footprint coverage was linked to outdated input imagery and other contextual factors (e.g., forest canopy coverage) affecting the automatic extraction[19]. In addition, building footprints included non-residential buildings, which were likely to be more prominent in urban settlements.

Despite the limited building footprint coverage, their use as model covariates enabled us to produce meaningful inferences. The covariate *Average Building Proximity* had a positive effect on population densities in rural settlements, suggesting a link between population density and settlement compactness. The same covariate had a non-significant effect in urban settlements, potentially because of the complex settlement structure[14] and the higher prevalence of non-residential buildings[15]. The covariate *Average Building Focal Counts* had a positive effect on population densities in urban settlements, confirming the association between population density and urban centrality[14]. However, we observed an opposite effect in rural settlements, potentially because of higher population densities of settlements with a larger prevalence of residential buildings. Lastly, the covariate *Average Building Area* had a consistent negative effect on population densities, highlighting the impact of large non-residential buildings and small overcrowded buildings on population densities[15].

Similar to existing modeling efforts, the model assumed that the population totals retrieved from the household surveys were observed without error and that no people lived outside of the area defined using the building footprints[5,6]. Observation error in the household surveys is likely to result in lower population totals because of inaccessible areas within the microcensus clusters. Additional observation error in the building footprints is expected to underestimate the population totals where the satellite imagery used for automatic extraction is outdated or obfuscated by contextual factors[19]. These sources of error are likely to be systematic and could be tackled in future studies by including a measurement error component in the model[6]. The model also assumed constant age and sex structures at the province level by potentially failing to capture sub-provincial variations within the credibility intervals; a limitation that will be addressed in future studies through a more complex hierarchical structure[12]. Lastly, in future studies, we will develop a more systematic assessment of the implementation of weighted-precision approaches, particularly in the uncertainty associated with sampling weights[24].

The proposed bottom-up model advances the state-of-the-art of bottom-up population modeling in countries with outdated census data. Research to further develop bottom-up population models is underway in the eastern part of the DRC and other countries of sub-Saharan Africa as part of the Geo-Referenced Infrastructure and Demographic Data for Development (GRID3) programme[25]. These modeling efforts are tailored to the country context and the available input data. Although the applications of bottom-up population modeling, conducted as part of the GRID3 programme, are primarily focused in sub-Saharan Africa, this approach has wider applicability and can support different steps of the census implementation, from planning (e.g., updating sampling frames) to implementation (e.g., support planning and logistics), from quality assessment (e.g., assess census coverage) to data usage (e.g., data anonymization)[8].

## Material and Methods

### Household surveys

We accessed geolocated household surveys involving the complete enumeration of 926 microcensus clusters of approximately three settled hectares in five provinces located in the western part of the DRC. The data was collected across two rounds of household surveys led by the UCLA-DRC Health Research and Training Program based at the University of California, Los Angeles Fielding School of Public Health and the Kinshasa School of Public Health (KSPH)[26]. The first round of surveys was carried out in 2017 in the provinces of Kinshasa, Kwango, Kwilu, and Mai-Ndombe using random sampling, while the second round was carried out in 2018 in the provinces of Kinshasa and Kongo Central using population-weighted sampling[23]. In both surveys, seed locations (i.e., 100m grid cells) were selected and cluster boundaries were manually delineated around these locations to include approximately three settled hectares with similar characteristics. We accessed the sampling weights for the second round of surveys and assessed their statistical distribution to identify outliers associated with uncertainties in the gridded population data used in the sampling[24]. To limit the effects of outliers resulting from the gridded population data used in the sampling, we truncated the sampling weights at the 90th percentile of the statistical distribution. We retrieved population totals for the clusters from the population counts recorded within each household and imputed population in households with a nonresponse based on the mean population per household within the same cluster. We also retrieved population totals for standardized age (i.e., under one-year-old, five-year groups from 1 to 80, and above 80 years old) and sex (i.e., male and female) groups[9] within each province by aggregating individual survey records.

**Building footprints**

We accessed building footprints automatically extracted by Ecopia.AI in 2019 using satellite imagery provided by Maxar Technologies between 2009 and 2019 within the DRC[27]. Morphological summary attributes based on these building footprints are openly available for the study region and throughout sub-Saharan Africa at a resolution of approximately 100m[19]. The building footprints provided the most accurate approximation of the spatial distribution of populations across the five provinces. However, their coverage varies according to the year of the satellite imagery used in the feature extraction and other contextual factors (e.g., forest canopy coverage) affecting the automatic delineation[19]. We used building footprints to derive morphological and topological attributes, such as area, perimeter, number of nodes, and distance to the nearest feature[9,14,15]. We summarized these attributes within the survey clusters and grid cells of approximately 100m comprising the five provinces using basic summary statistics, such as the sum, mean, and coefficient of variation. We also produced the same summary statistics for focal windows of approximately 500m, 1km, and 2km to reflect contextual characteristics[14]. We allocated the survey clusters and the grid cells to urban and rural settlements using an existing morphological classification derived from the same building footprint data[15,28]. We labeled the original *built-up area* class as urban settlement and merged the original classes *small settlement area* (i.e., representing rural settlements) and *hamlet* (i.e., isolated rural settlements) into a class labeled rural settlement.

**Administrative boundaries**

We accessed administrative boundaries provided by the Bureau Central du Recensement (BCR), the administrative body responsible for the census implementation in the DRC[29]. The boundaries comprised the administrative level 0 (i.e., country), level 1 (i.e., provinces), level 2 (i.e., territories and cities), and level 3 (i.e., sectors/chiefdoms and municipalities). At the time of this study, the administrative boundaries were being consolidated, and level 3 boundaries were only available for the city of Kinshasa. We first derived the spatial extent of the provinces from the level 1 boundaries. We then created the sub-provincial regions by combining level 2 and level 3 boundaries. In doing so, we merged the 24 municipalities comprising the city of Kinshasa into nine contiguous groups of municipalities with similar land use characteristics[14] to prevent the presence of unsurveyed sub-provincial units. We produced gridded datasets with a resolution of approximately 100m with unique identifiers for each province and sub-provincial region and subsequently allocated the survey clusters to a single province and a sub-provincial region.

**Covariate processing and selection**

We constrained the extent of the clusters using the building footprints located within a radius of approximately 50m from the surveyed households to exclude unsurveyed areas associated with accessibility constraints. We derived morphological and topological attribute summaries from the building footprints and extracted additional summaries from standard gridded datasets used in the study of population distributions[30]. We visually assessed scatterplots and Pearson correlations between log-population densities (people/building footprint ha) and the attribute summaries across the clusters to select model covariates. We retained the five covariates with the highest correlation coefficient — *Building Count* (count of structures), *Average Building Area* (in ha), *Average Building Perimeter* (in m), *Average*

*Building Proximity* or the inverse of the distance to the nearest building (in m), and *Average Building Focal Count* (average count of building within a focal window of approximately 2km). We assessed Pearson correlations between the five covariates and subsequently discarded *Building Count* and *Average Building Perimeter* as strongly correlated with other covariates to avoid multicollinearity. The selected covariates were finally scaled based on the mean and standard deviation computed at the grid-cell level across the study area.

Data processing and covariate selection were conducted in R version 4.0.2[31] using the R packages *raster*[32] and *sf*[33].

**Population model**

We modeled population totals by extending an existing hierarchical Bayesian modeling framework for population estimation[5]. Eq. (**1**) models the total number of people $N_i$ as a Poisson process, where $D_i$ is the population density (people/ building footprint ha) and $A_i$ is the total area of building footprints(ha) derived from the building footprints within each microcensus cluster $i$.

$$N_i \sim Poisson(D_i\, A_i) \tag{1}$$

Eq. (**2**) models $D_i$ as a log-normal process to relax the assumptions of the Poisson distribution, where $\bar{D}_i$ is the expected population density on a log-scale and $\tau_{t,p,i}$ is a hierarchical precision term estimated by settlement type $t$ and province $p$ for each cluster $i$.

$$D_i \sim LogNormal(\bar{D}_i, \tau_{t,p,i}) \tag{2}$$

Eq. (**3**) defines the precision term $\tau_{t,p,i}$ based on a hierarchical estimate of precision $\tau_{t,p}$ and the model weights $v_i$, similar to an existing simulation[11]. $\tau_{t,p}$ is estimated hierarchically by settlement type $t$ and province $p$ using uninformative priors on the mean $\mu_{t,p}$ and the variance $\sigma_{t,p}$ terms, which are modeled by a normal and uniform distribution, respectively.

$$\tau_{t,p,i} = \sqrt{\frac{1}{v_i \tau_{t,p}^{-2}}} \tag{3}$$

$$\tau_{t,p} \sim Half\text{-}Normal(\mu_{t,p}, \sigma_{t,p})$$

$$\mu_{t,p} \sim Half\text{-}Normal(\mu_t, \sigma_t)$$

$$\sigma_{t,p} \sim Uniform(0, \sigma_t)$$

$$\mu_t \sim Normal(0, 1000)$$

$$\sigma_t \sim Uniform(0, 1000)$$

Eq. (**4**) defines the model weight $v_i$ as the inverse of the sampling weight $w_i$ used to select the cluster $i$ in the second round of household surveys. The sum of $w_i$ is used to proportionally impute $w_i$ for the clusters that were selected randomly during the first round of household surveys. $v_i$ are then rescaled to sum to one across all the clusters $I$.

$$v_i = \frac{w_i^{-1}}{\sum_{i=1}^{I} w_i^{-1}} \tag{4}$$

As the estimate of precision $\tau_{t,p,i}$ cannot be derived in locations where the model weights $w_i$ are not available and adopted for posterior model predictions, Eq. (**5**) determines a hierarchical estimate of precision $\hat{\tau}_{t,p}$ from a weighted average of $\tau_{t,p,i}$, where $I_{t,p}$ is the number of clusters $i$ within settlement type $t$ and province $p$.

$$\hat{\tau}_{t,p} = \frac{\sum_{i=1}^{I_{t,p}} \tau_{t,p,i} \sqrt{v_i}}{\sum_{i=1}^{I_{t,p}} \sqrt{v_i}} \tag{5}$$

Eq. (**6**) uses the precision estimate $\hat{\tau}_{t,p}$ for posterior model predictions by altering Eq. (**2**).

$$\widehat{D}_i \sim LogNormal(\overline{D}_i, \hat{\tau}_{t,p}) \tag{6}$$

Eq. (**7**) models the expected population density $\overline{D}_i$ using a linear regression with random intercept $\alpha_{t,p,l}$ estimated by settlement type $t$, province $p$ and local area $l$, and $K$ covariates $x_k$ with random effects $\beta_{k,t}$ estimated by settlement type $t$.

$$\overline{D}_i = \alpha_{t,p,l} + \sum_{k=1}^{K} \beta_{k,t} \, x_{k,i} \tag{7}$$

Eq. (**8**) models the hierarchical intercept $\alpha_{t,p,l}$ for a local area $l$ belonging to a settlement type $t$ and province $p$ as a nested hierarchy with uninformative priors on the mean $\xi_{t,p}$ and variance $v_{t,p}$ terms. These are modeled using a normal and uniform distribution, respectively.

$$\alpha_{t,p,l} \sim Normal(\xi_{t,p}, v_{t,p}) \tag{8}$$

$$\xi_{t,p} \sim Normal(\xi_t, v_t)$$

$$v_{t,p} \sim Uniform(0, v_t)$$

$$\xi_t \sim Normal(0, 1000)$$

$$v_t \sim Uniform(0, 1000)$$

Eq. (**9**) models the random effects $\beta_{k,t}$ for each covariate $k$ independently for each settlement type $t$ with uninformative priors on the mean $\rho_k$ and variance $\omega_k$ terms, which follow a normal and uniform distribution, respectively.

$$\beta_{k,t} \sim Normal(\rho_k, \omega_k) \tag{9}$$

$$\rho_k \sim Normal(0, 1000)$$

$$\omega_k \sim Uniform(0, 1000)$$

For each covariate $k$, random effects $\beta_{k,t}$ with similar estimated posterior distributions across settlement types $t$ are converted into a fixed effect $\beta_k$ modeled with an uninformative normal distribution (Eq. (**10**)).

$$\beta_k \sim Normal(0, 1000) \tag{10}$$

**Age and sex structure model**

Eq. (**11**) models age and sex structures at the province level as the proportion of the population $\pi_{g,p}$ derived from the observed total population $N_p$ within the province $p$ and the relative age and sex breakdowns $N_{g,p}$ within $G$ demographic groups $g$. $g$ consists of two sex groups (i.e., male and female) subdivided into 18 age groups (i.e., under one-year-old, one to four years old, five-year groups from five to 80, and above 80 years old).

$$N_{g,p} \sim Multinomial(N_p, \pi_{g,p}) \tag{11}$$

Eq. (**12**) uses an uninformative Dirichlet distribution[13] as a conjugate prior for $\pi_{g,p}$ where $\chi^G$ is a constant numerical vector with values $1/G$ and of length $G$.

$$\pi_{g,p} \sim Dirichlet(\chi^G) \tag{12}$$

**Model fit and diagnostics**

We estimated the model with Markov chain Monte Carlo (MCMC) methods in JAGS[34] using the R package runjags[35]. We assessed convergence of three MCMC chains using the Gelman-Rubin statistic, and values less than 1.1 were interpreted as indicating convergence[36]. We tested the model residuals for spatial autocorrelation using semivariograms and Moran's I statistics. We examined model fit in- and out-of-sample using 10-fold cross-validation, where the model was fit ten times, each time withholding a random 10% of survey clusters until all had been held out once. To assess model fit for age and sex proportions, we held out 10% of the clusters for each province and assessed the combined posterior distribution for each demographic group. For in- and out-of-sample predicted population sizes, densities, and province-level age and sex proportions, we evaluated bias (i.e., the mean of residuals — mean posterior predictions minus observed values), imprecision (i.e., the standard deviation of residuals), inaccuracy (i.e., the mean of absolute residuals), $R^2$ values (i.e., the squared Pearson correlation coefficient among the residuals), and the percentage of observations falling within the 95% prediction intervals[36]. We also computed bias, imprecision, and inaccuracy using standardized residuals (i.e., residuals divided by the mean posterior predictions).

**Data Availability**

Data supporting this study have been deposited on Figshare with the accession codes ### (input data stored as an .RData file) and ### (jags model as an .R file). The model results can be accessed on a web repository[16], visualized using a web-mapping application[17], and implemented in data analyses through an R package[18].

## Acknowledgments

This work is part of the GRID3 project (Geo-Referenced Infrastructure and Demographic Data for Development), funded by the Bill and Melinda Gates Foundation and the United Kingdom Foreign, Commonwealth and Development Office (FCDO) (#INV009579). Project partners include WorldPop at the University of Southampton, the United Nations Population Fund (UNFPA), the Center for International Earth Science Information Network (CIESIN) in the Earth Institute at Columbia University, and the Flowminder Foundation.

The UCLA-DRC Health Research and Training Program based at the University of California, Los Angeles Fielding School of Public Health, the Kinshasa School of Public Health (KSPH) led the two rounds of household surveys in 2017 and 2018, with the support of the Bureau Central du Recensement (BCR) (#OPP1151786). Prof Emile Okitolonda-Wemakoy at the KSPH, who passed away before the submission of this work, provided oversight to the household survey data collection. D'Andre Spencer, Camille Dzogang, Jojo Mwanza, Handdy Kalunga, Millet Mfawankang, Eric Musenge, Elie Lokutumba, Joseph Wasiswa, Arthur Lisambo, Kevin Karume, Kizito Mosema, Lievin Dinoka, and Michael Beya supervised the surveyors who were hired locally in collaboration with the provincial health departments.

The Oak Ridge National Laboratory (ORNL) supported the first round of household surveys. Key ORNL collaborators include Eric Weber, Jeanette Weaver, and St. Thomas LeDoux. The survey data collection instrument and data quality control platform were developed by eHealth Africa in collaboration with the UCLA-DRC program. Key eHealth Africa collaborators include Ayodele Adeyemo, Dami Sonoiki, and Adeoluwa Akande. The health zone bureau staff throughout the five provinces provided logistical support to surveyors as they travelled to microcensus clusters within each health zone. We acknowledge the work of local surveyors who carried out the survey data collection, often in the face of significant logistical challenges in remote, difficult-to-traverse areas.

The authors used the IRIDIS High-Performance Computing Facility and associated support services at the University of Southampton.

## Author Contributions

G.B. prepared the manuscript; E.D., D.L.R., C.A.D., H.R.C., A.N.L., K.T., C.S., N.A.H., T.F., K.M., A.B., A.W.R., and A.J.T. edited the manuscript; G.B., E.D., H.R.C., K.T., C.S., N.A.H., T.F., K.M., and A.B. supervised data collection; G.B., E.D, and H.R.C. processed the data; G.B., E.D., D.L.R., and C.A.D. developed the model; G.B., E.D., and D.R.L. implemented the model; H.R.C., A.N.L., K.T., and A.J.T. provided project oversight; A.W.R. and A.J.T. acquired funding.

## Competing Interests

The authors declare no competing interests.

## Figure Legends and Tables

**Fig 1.** Estimated population totals per approximately 100m grid cell in the capital cities of the provinces of (**A**) Kinshasa (Kinshasa), (**B**) Kongo Central (Matadi), (**C**) Kwango (Kenge), (**D**) Kwilu (Bandundu), and (**E**) Mai-Ndombe (Inongo). The map of the DRC shows the extent of the five provinces in grey.

**Fig 2.** Observed population densities (people/building footprint ha) in the 905 microcensus clusters according to settlement type for the provinces of (**A**) Kinshasa — 25 rural and 254 urban clusters, (**B**) Kongo Central — 113 rural and 108 urban clusters, (**C**) Kwango — 110 rural and 6 urban clusters, (**D**) Kwilu — 183 rural and 23 urban clusters, and (**E**) Mai-Ndombe — 69 rural and 12 urban clusters.

**Fig 3.** Posterior probability distributions of the random intercepts (people/building footprint ha) by settlement type across the provinces of (**A**) Kinshasa, (**B**) Kongo Central, (**C**) Kwango, (**D**) Kwilu, and (**E**) Mai-Ndombe. The black dots show the mean of the distributions, while the horizontal black lines show the 95% credible intervals.

**Fig 4.** Posterior probability distribution of the random effect by settlement type (i.e., urban and rural) for the covariates *Average Building Proximity* and *Average Building Focal Count* and the fixed effect for the covariate *Average Building Area*. The black dots show the mean of the distributions, while the horizontal black lines show the 95% credible intervals.

**Fig 5.** Means of the posterior distribution of age and sex proportions for the provinces of (**A**) Kinshasa, (**B**) Kongo Central, (**C**) Kwango, (**D**) Kwilu, and (**E**) Mai-Ndombe. The horizontal black lines show the 95% credible intervals.

**Fig 6.** Observed microcensus-cluster level totals (people) and densities (people/building footprint ha) versus in-sample and out-of-sample mean posterior predictions (dots) with 95% credible intervals (vertical lines). Population totals and densities are classified by settlement type (i.e., urban and rural). The diagonal black lines show a perfect relationship between observations and predictions.

**Table 1.** Analysis of residuals for the estimated population totals (people), population densities (people/building footprint ha), and province-level age and sex proportions for in-sample and out-of-sample posterior predictions. Bias represents the mean of the residuals, imprecision the standard deviation of residuals, inaccuracy the mean of absolute residuals, $R^2$ the squared Pearson correlation coefficient among the residuals, and the percentage of observations falling within the 95% credible intervals. Values in parentheses are computed using scaled residuals (residuals/predictions).